\documentclass[12pt,a4paper]{article}

\usepackage{amssymb,amsfonts}
\usepackage{braket}
\usepackage[centertags]{amsmath}
\usepackage[dvips]{graphicx}
\usepackage{psfrag}
\usepackage{color}
\usepackage{bbm}
\newcommand{\be}{\begin{equation}}
\newcommand{\ee}{\end{equation}}
\newcommand{\bea}{\begin{eqnarray}}
\newcommand{\eea}{\end{eqnarray}}
\newcommand{\bdm}{\begin{displaymath}}
\newcommand{\edm}{\end{displaymath}}
\newcommand{\lb}{\label}
\newcommand{\I}{\mbox{i}}
\newcommand{\D}{\mbox{d}}
\newcommand{\E}{\mbox{e}}

\begin{document}
\begin{titlepage}

\begin{center}
{\large\bf  INTERPRETATION OF THE TRIAD ORIENTATIONS IN
LOOP QUANTUM COSMOLOGY}
\vskip 1cm
{\bf Claus Kiefer and Christian Schell\footnote{Address from October
  1, 2012: Max Planck Institute for Gravitational Physics (Albert
  Einstein Institute), Am M\"uhlenberg~1, 14476 Potsdam, Germany.}}
\vskip 0.4cm
 Institut f\"ur Theoretische Physik, Universit\"at zu K\"oln,\\
 Z\"ulpicher Stra\ss e~77, 50937 K\"oln, Germany.
\end{center}
\date{\today}
\vskip 2cm
\begin{center}
{\bf Abstract}
\end{center}
\small
\begin{quote}
Loop quantum cosmology allows for arbitrary superpositions of the
triad variable. We show here how these superpositions can become
indistinguishable from a classical mixture by the interaction with
fermions. We calculate the reduced density
matrix for a locally rotationally symmetric Bianchi I model and show
that the purity factor for the triads decreases by decoherence.
In this way, the universe assumes a definite orientation.
\end{quote}
\normalsize
\end{titlepage}

\section{Introduction}

Quantum theory seems to be a universal framework for
all interactions. As such, it should be applicable to the universe
as a whole, leading to quantum cosmology. Since gravity is the
dominant interaction on large scales, the formalism of quantum
cosmology must be based on a theory of quantum gravity. 

A direct quantization of general relativity in the canonical formalism
leads to quantum geometrodynamics and the Wheeler--DeWitt equation
\cite{OUP}. An
alternative canonical approach is loop quantum gravity \cite{OUP,Thiemann},
which leads to loop quantum cosmology when applied to cosmological models 
\cite{Bojowald,AP11}. In loop quantum cosmology, the quantum state obeys a
difference equation and not a differential equation.
The reason for this is the presence of discrete spectra for geometric
operators (such as the area operator) in the full theory. 
This discreteness seems to facilitate the avoidance of the classical big-bang
singularity. Singularity avoidance thus seems to be more generic than
in Wheeler--DeWitt quantum cosmology, although in the latter there
exist models where the classical singularities are avoided, too
\cite{BKSM09}.  

In loop quantum cosmology, one does not work with 
three-dimensional metrics, but with triads. This leads to an important
difference to quantum geometrodynamics: triads can have two
different orientations (left-handed and right-handed), which both lead
to the same metric. A central question then concerns the role of the
two orientations in the quantum theory. It has been claimed, for
example, that the orientation of the triad reverses in a big-bang
transition where the universe turns it inside out (\cite{Bojowald},
p.~67). This assumes that the universe is always in a state of
definite orientation. But because of its linear
structure, quantum cosmology allows the occurrence of arbitrary
superpositions of the two orientations; one would then expect that the
universe is unlikely to be found in a definite orientation and that
it is meaningless to talk about a change of orientation. How can
this be understood?
 
This situation reminds one of an old problem for chiral molecules posed by Hund
in the 1920s \cite{Hund}. Many molecules occur as objects with a
definite shape, although the underlying Hamiltonian is
parity-invariant. Energy eigenstates, which are superpositions of
chiral states, are typically found only for
small molecules, such as ammonia, but not for bigger molecules such as
sugar. The reason for the occurrence of chiral states as robust
quantities was understood in the last decades and can be traced back
to the process of decoherence -- the irreversible emergence of
classical properties by the unavoidable interaction with the
environment \cite{deco,Schlosshauer}. The spatial orientation of
molecules is fixed by the scattering with photons and air molecules;
the information about the superposition between different chiral
states is transferred into the entanglement between the chiral molecules and
the photons or air molecules and is no longer accessible at the
molecules themselves. 

In this paper, we shall examine whether decoherence is also
responsible for the appearance of definite triad orientations in loop
quantum cosmology. For this purpose, we need degrees of freedom that
may serve as an appropriate `environment' for the triads, in analogy
to the photons or air molecules for the chiral molecules. Such an
environment must be of fermionic nature, because bosonic variables 
only interact with the metric and are unable to discriminate between the
two different orientations of a triad. (We assume here that there is
no parity violation in the matter coupling.)

One possibility to implement fermions into quantum cosmology is to consider
inhomogeneous fermionic fluctuations superimposed on a homogeneous
background \cite{DEH87}. In fact, decoherence by fermions in quantum
cosmology was discussed at length some time ago \cite{CK89,BKK99}.
It was found there that a fermionic environment leads to a
suppression of interferences for the scale factor and a background
scalar field and thus to their classical appearance, but in a way that is
less efficient than for a bosonic environment; the reason for this smaller
efficiency can be traced back to the Pauli principle. Inhomogeneous (bosonic)
fluctuations can, of course, become themselves relevant as a quantum
system subject to decoherence. This is the case for the primordial
fluctuations in inflationary cosmology, which can serve as the seeds
for the origin of structure in the universe; their decoherence is
discussed in \cite{KLPS07} and other papers. 

Here, we shall follow another route. Instead of taking inhomogeneous
fermionic fluctuations, we retain homogeneity, but relax
isotropy. Homogeneity demands any fields, including fermions, to be
spatially constant. Imposing, in addition, isotropy would have the
consequence that for fermions this constant is zero, because otherwise
one could construct from the spinor fields a non-zero vector with a
given direction, in contradiction to isotropy \cite{BD08}. 
We therefore consider one of the simplest possible anisotropic models
-- the Bianchi I model (see, for example, \cite{RS75} for a general
introduction into classical and quantum aspects for such models).
It is, in fact, the fermionic currents that in the considered model
lead to the anisotropies.
To simplify matters, we reduce this model further by imposing a
rotational symmetry with respect to {\em one} axis, that is, by
introducing the additional isotropy group U(1);
one then arrives at a locally rotationally symmetric (LRS) model
in which two of the diagonal components of the connection and the
triad are equal. This was also done in \cite{BD08}. The fermionic
current aligned in the 1-direction of internal space is then the sole
reason for the anisotropy.

This model is admittedly very simple. We do, however, not expect that
any of the qualitative features discussed here will change when going to
more complicated models (such as the Gowdy model), although the technical
difficulties will considerably increase.

The environment for the triads in the sense of decoherence is thus a
homogeneous fermion field. It is known from quantum mechanics that one
does not necessarily need many degrees of freedom for decoherence. For
example, one can get decoherence from a long-range
(e.g. gravitational) interaction between two bodies, where one body
decoheres the other (\cite{JZ85}, p.~228).  

In this paper, we shall basically use the model presented in
\cite{BD08} (fermions in a LRS Bianchi I model) and integrate over the
fermions to study their decohering influence on the triads. But in order
to investigate decoherence as a dynamical process, we have to
introduce some notion of time. Quantum gravity is fundamentally
timeless in the sense that there is no external time parameter present
\cite{OUP}. We thus introduce an inner variable and choose a
massless scalar field $\varphi$ for this purpose. Such a field was
used as an inner time in loop quantum cosmology for the Friedmann
case in \cite{APS06} and for the Bianchi I case (without fermions) in 
\cite{AWE09}. 

Section~2 contains a brief review of the formalism. Section~3 is the
main part of our paper. We calculate the reduced density matrix and
the linear entropy by tracing out the fermions and discuss the
resulting decoherence. Section~4 contains our conclusions.

\section{The formal framework}

The formal framework of our discussion
is based on the model of \cite{BD08} supplemented by a massless scalar field. 
For the general formalism of fermions in loop quantum gravity, we
refer to \cite{fermions} and the references therein. As for the units,
we set $c=1$ but keep $G$ and $\hbar$ (although $G$ will not appear
explicitly in \eqref{differenceeqn} below).

In loop quantum gravity, the fundamental variables are constructed
from the
Ashtekar--Barbero connection $A^i_a(x)$ and the triad field $E^a_i(x)$
\cite{OUP,Bojowald}. Here, $i=1,2,3$ is the internal index of the
$\mathfrak{su}(2)$ algebra, and $a=1,2,3$ is the usual space
index. The fundamental 
variables are fluxes (integrals of the $E^a_i(x)$ over two-dimensional
surfaces) and holonomies (integrals of the $A^i_a(x)$ along
curves). For Friedmann--Lema\^{\i}tre cosmology, the connection and
the triad reduce to $A^i_a=c\delta^i_a$ and $E^a_i=p\delta^a_i$,
respectively, where $\vert p\vert=a^2$ ($a$ is the scale factor), and $c$ is at
the classical level given by $c=\gamma\dot{a}$, where $\gamma$ denotes
the Barbero--Immirzi parameter.\footnote{The Barbero--Immirzi
  parameter is called $\beta$ in \cite{OUP}, but we stick here to the
  convention of \cite{BD08}.}

In the Bianchi I model, we have three scale factors
instead of one. Therefore, instead of $c$ we have three variables
called $c_1$, $\tilde{c_2}$, and $\tilde{c_3}$ (we use here
tildes to follow the convention of \cite{BD08}); instead of $p$, we
have $p^1$, $p^2$, and $p^3$. In the LRS Bianchi I model used
here, we have $\tilde{c_2}=\tilde{c_3}$ and $p^2=p^3$.

The fermions are described by a densitized current
${\mathcal J}_{\alpha}$, $\alpha=0,1,2,3$; we assume here, for
simplicity, that ${\mathcal J}_2={\mathcal J}_3=0$. The fermionic
current is a source of torsion, which leads to a new variable
denoted below by $\phi$;
this variable is, however, not a physical degree of freedom and can be connected
to the fermionic current by the Gauss constraint.
The fermionic current depends on the four half-densitized 
Grassmann variables $\Theta_1$, $\Theta_2$, $\Theta_3$, and $\Theta_4$
(summarized below under the label $\Theta$). In the quantum theory,
one has for the non-vanishing
components of the fermionic current operator the expressions
\be
\hat{\mathcal{J}}_0/\hbar=
\partial_{\Theta_1}\Theta_1+\partial_{\Theta_2}\Theta_2-\partial_{\Theta_3} 
\Theta_3-\partial_{\Theta_4}\Theta_4
\ee
and
\be
 \hat{\mathcal{J}}_1/\hbar=\partial_{\Theta_2}\Theta_1+
\partial_{\Theta_1}\Theta_2+
\partial_{\Theta_4}\Theta_3+\partial_{\Theta_3}\Theta_4. 
\ee

In addition to the variables employed in \cite{BD08}, we introduce here a
massless scalar field, which serves as an inner time variable. 
The classical Hamiltonian for it reads 
\be
H_{\varphi}=8\pi
Gp_{\varphi}^2(\sqrt{|p^1|}|p^2|)^{-1}. 
\ee
(Because this Hamiltonian depends only on the absolute values of $p^1$ and
$p^2$, one recognizes that it is insensitive to the triad orientation;
this is why 
non-fermionic fields are not suitable to serve as a decohering environment.)
For its quantization, we use the same technique as in \cite{BD08}. 
The action of $\hat{p}_{\varphi}$ on a state is given by
$-\I\hbar\partial_{\varphi}$. 

Since one of the fundamental variables in loop quantum gravity is the
holonomy, the connection has to be exponentiated also in loop
quantum cosmology in order to arrive at mathematically well defined
expressions. (In homogeneous situations, the densitized triad
components can be directly promoted to operators.) Functions on 
configuration space can then be expanded as follows:
\be
g(c_1,\tilde{c_2},\phi)=\sum_{\mu_1,\mu_2,k}
\xi_{\mu_1,\mu_2,k}\exp\left(\frac12\I\mu_1c_1+
\frac12\I\mu_2\tilde{c_2}+\I k\phi\right),
\ee
where the sum is over finitely many $\mu_1,\mu_2\in {\mathbb R}$, and
$k\in {\mathbb Z}$. By solving the
Gauss constraint for $p_{\phi}$ (the variable conjugate to $\phi$), we
arrive, as in \cite{BD08}, 
at the difference equation \eqref{differenceeqn} below, in which $k$
is eliminated. Therefore, we drop $k$ in the following.  

A general state $\ket{s}$ is defined in the full Hilbert space 
${\mathcal H}= {\mathcal H}_{\rm grav}\otimes {\mathcal 
  H}_{\rm fermion}\otimes {\mathcal H}_{\varphi}$ and can be expanded as
\be 
\ket{s}=
\sum_{\mu_1,\mu_2}s(\mu_1,\mu_2,\Theta,\varphi)\ket{\mu_1,\mu_2}\label{state}, 
\ee
where the $\ket{\mu_1,\mu_2}$ denote the common eigenstates of the triad
operators $\hat{p}^1$ and $\hat{p}^2$. Since $\mu_2\to -\mu_2$
corresponds to a triad rotation, we demand
\be
s(\mu_1,\mu_2,\Theta,\varphi)=
s(\mu_1,-\mu_2,\Theta,\varphi).
\ee
In contrast to $\mu_2$, the sign of $\mu_1$ determines the relative
orientation of the triad.

For the symmetric factor ordering, the difference equation for the
total quantum state including
the massless scalar field then reads\footnote{For simplicity, 
we choose $\alpha\rightarrow\infty$ for the non-minimal coupling
parameter $\alpha$ in \cite{BD08}, which also leads to $\theta=1$ and
$\beta=\gamma$ for the parameters $\theta$ and $\beta$ occurring in 
\cite{BD08}. This has no qualitative influence on
our results.} 

\begin{eqnarray}
\hat{h}s(\mu_{1},\mu_{2},\Theta,\varphi):= 
2\left[(|\mu_{2}+3\delta_{2}|-|\mu_{2}+\delta_{2}|)\sqrt{|\mu_{1}+2\delta_{1}|}\right.\nonumber  
\\ 
\left.+(|\mu_{2}+\delta_{2}|-|\mu_{2}-\delta_{2}|)\sqrt{|\mu_{1}|}\right]
s(\mu_{1}+2\delta_{1},\mu_{2}+2\delta_{2},\Theta,\varphi)\nonumber \\
-2\left[(|\mu_{2}+3\delta_{2}|-|\mu_{2}+\delta_{2}|)\sqrt{|\mu_{1}-2\delta_{1}|}\right.\nonumber
\\ 
\left.+(|\mu_{2}+\delta_{2}|-|\mu_{2}-\delta_{2}|)\sqrt{|\mu_{1}|}\right]
s(\mu_{1}-2\delta_{1},\mu_{2}+2\delta_{2},\Theta,\varphi)\nonumber \\
+2\left[(|\mu_{2}-\delta_{2}|-|\mu_{2}-3\delta_{2}|)\sqrt{|\mu_{1}-2\delta_{1}|}\right.\nonumber
\\ 
\left.+(|\mu_{2}+\delta_{2}|-|\mu_{2}-\delta_{2}|)\sqrt{|\mu_{1}|}\right]
s(\mu_{1}-2\delta_{1},\mu_{2}-2\delta_{2},\Theta,\varphi)\nonumber \\
-2\left((|\mu_{2}-\delta_{2}|-|\mu_{2}-3\delta_{2}|)\sqrt{|\mu_{1}+2\delta_{1}|}\right.\nonumber
\\ 
\left.+(|\mu_{2}+\delta_{2}|-|\mu_{2}-\delta_{2}|)\sqrt{|\mu_{1}|}\right]
s(\mu_{1}+2\delta_{1},\mu_{2}-2\delta_{2},\Theta,\varphi))\nonumber \\
+\left(\sqrt{|\mu_{1}+\delta_{1}|}-\sqrt{|\mu_{1}-\delta_{1}|}\right)\nonumber \\
\cdot\left[(|\mu_{2}|+|\mu_{2}+4\delta_{2}|)s(\mu_{1},\mu_{2}+4\delta_{2},\Theta,\varphi)\right.\nonumber
\\ 
\left.-4|\mu_{2}|s(\mu_{1},\mu_{2},\Theta,\varphi)+(|\mu_{2}|+|\mu_{2}-4\delta_{2}|)s(\mu_{1},\mu_{2}-4\delta_{2},\Theta,\varphi)\right]\nonumber \\
=-\frac{27}{4}|\mu_{1}|^{1/3}|\mu_{2}|^{1/3}(|\mu_{1}+\delta_{1}|^{1/6}-|\mu_{1}-\delta_{1}|^{1/6})\nonumber
\\ 
(|\mu_{2}+\delta_{2}|^{1/3}-|\mu_{2}-\delta_{2}|^{1/3})^{2}\nonumber
\\ \times
\left[{\left(1+4\gamma^{2}-\frac{2\gamma^{2}}{1+\gamma^{2}}\left(3+2\gamma^{2}\right)-\frac{1}{1+\gamma^{2}}\right)}\frac{\hat{\mathcal{J}}_{1}^{2}}{\hbar^{2}}\right.\nonumber
\\ 
\left.+{\frac{3\gamma^{2}}{1+\gamma^{2}}}\frac{\hat{\mathcal{J}}_{0}^{2}}{\hbar^{2}}-16\frac{\hat{p}_{\varphi}^{2}}{\hbar^{2}}
\right]s(\mu_{1},\mu_{2},\Theta,\varphi).\label{differenceeqn}\end{eqnarray}
The increments $\delta_1$ and $\delta_2$ appearing in this difference
equation arise from the edge lengths of spin networks in the full
theory; in minisuperspace, their exact form and meaning remains open.

The gravitational constant $G$ does not appear explicitly here because
it has been absorbed in the process of expressing the inverse volume
in terms of Poisson brackets \cite{BD08}.

We can write the difference equation \eqref{differenceeqn} in the form
\be
\lb{h}
\hat{h}s(\mu_{1},\mu_{2},\Theta,\varphi)=
-\frac{27{\mathcal T}}{4}
\left(c_{1}\frac{\hat{\mathcal{J}}_{1}^{2}}{\hbar^{2}}+
c_{0}\frac{\hat{\mathcal{J}}_{0}^{2}}{\hbar^{2}}+
c_{\varphi}\frac{\widehat{p_{\varphi}}^{2}}{\hbar^{2}}\right) 
s(\mu_{1},\mu_{2},\Theta,\varphi),
\ee
where we have introduced 
\be
{\mathcal T}:=
|\mu_{1}|^{1/3}|\mu_{2}|^{1/3}(|\mu_{1}+\delta_{1}|^{1/6}-
|\mu_{1}-\delta_{1}|^{1/6})
(|\mu_{2}+\delta_{2}|^{1/3}-|\mu_{2}-\delta_{2}|^{1/3})^{2} 
\ee
and 
\be
c_{1}  := 
-\frac{\gamma^2}{1+\gamma^2},\quad
c_{0}  := {\frac{3\gamma^{2}}{1+\gamma^{2}}},\quad
c_{\varphi}  :=  -16.\ee
We shall now discuss how superpositions of different triad
orientations can be suppressed by the interaction with fermions.

\section{Decoherence of triad orientations}

In the following, we shall make
the simplifications $\Theta_2=\Theta_3=\Theta_4=0$, $\Theta_1\equiv\Theta $;
we thus have $\hat{\mathcal{J}}_{1}^{2}=0$ and 
$\hat{\mathcal{J}}_{0}^2/\hbar^2=\hat{\mathcal{J}}_{0}/\hbar
=\partial_{\Theta}\Theta=\mathbbm{1}-\Theta\partial_{\Theta}$. 
The full equation \eqref{h} can then be written in the form
\be
\lb{gs}
\frac{27{\mathcal T}}{4}c_0\Theta\partial_{\Theta}s(\mu_1,\mu_2,\Theta,\varphi)=
\hat{g}s(\mu_1,\mu_2,\Theta,\varphi), 
\ee 
with
\be
\hat{g}:=\hat{h}+\frac{27{\mathcal
    T}}{4}\left(c_0+c_{\varphi}\frac{\hat{p}_{\varphi}^2}{\hbar^2}\right).
\ee
We can then make the following ansatz for the solution:
\be
\lb{ansatz}
s(\mu_1,\mu_2,\Theta,\varphi)=s_0(\mu_1,\mu_2,\varphi)+\Theta
s_1(\mu_1,\mu_2,\varphi). 
\ee
 From \eqref{gs}, one then gets the following two equations:
\bea
\hat{g}s_1(\mu_1,\mu_2,\varphi) &=& \frac{27{\mathcal T}}{4}c_0
s_1(\mu_1,\mu_2,\varphi),\\
\hat{g}s_0(\mu_1,\mu_2,\varphi) &=& 0.\label{s0ands1}
\eea
Since the total system is in a pure state, the total density matrix reads
\be
\lb{rhototal}
\rho_{\rm tot}=\bar{s}(\mu_1',\mu_2',\Theta',\varphi)
               s(\mu_1,\mu_2,\Theta,\varphi).
\ee
Using the rules for Grassmann integration (see e.g. \cite{FS}), one can define
the following reduced density matrix for the gravitational variables
alone:
\be
\rho_{\rm red}(\mu_1,\mu_2;\mu_1',\mu_2';\varphi)=\int\D\Theta\D\bar{\Theta}\
\E^{-\Theta\bar{\Theta}} 
\bar{s}(\mu_1',\mu_2',\Theta,\varphi)
               s(\mu_1,\mu_2,\Theta,\varphi).
\ee
The integration yields
\bea
\lb{rhored}
& & \rho_{\rm
  red}(\mu_1,\mu_2;\mu_1',\mu_2';\varphi)=\nonumber\\
& & \;\;\bar{s_1}(\mu_1',\mu_2',\varphi) 
s_1(\mu_1,\mu_2,\varphi)+\bar{s_0}(\mu_1',\mu_2',\varphi)
s_0(\mu_1,\mu_2,\varphi).
\eea
The total density matrix corresponds to an entangled state if 
both terms are present
in the expression \eqref{rhored} for the
reduced density matrix, that is, if both $s_0$
and $s_1$ are non-vanishing. Otherwise, 
the gravitational part is by itself in a pure state and there is no
decoherence. This is also obvious from \eqref{ansatz}. 

A measure for the purity of the total state \eqref{rhototal} is
the trace of $\rho_{\rm red}^2$, which is equal to one for a pure state and
smaller than one for a mixed state; it is directly related to the
linear entropy $S_{\rm lin}=1-{\rm tr}\rho_{\rm red}^2$
\cite{deco}. One could also address the von Neumann entropy
$-k_{\rm B}{\rm tr}\left(\rho_{\rm red}\ln\rho_{\rm red}\right)$, but for the
present purpose it is sufficient to restrict to $S_{\rm lin}$.

To obtain concrete results, we calculate the evolution of the
coefficients $s_0$ and $s_1$ in \eqref{s0ands1}.  
The spatial parameters $\mu_1$ and $\mu_2$ span a two-dimensional
discrete lattice on which we have to choose appropriate initial data.  
Solving the equations in \eqref{s0ands1} for the derivative in the
continuous variable $\varphi$, we have 
\begin{eqnarray}
 \partial_{\varphi}^2s_1=-\frac{1}{108}\frac{1}{{\mathcal T}}\hat{h}s_1,\\
 \partial_{\varphi}^2s_0=-\frac{1}{108}\frac{1}{{\mathcal
     T}}\hat{h}s_0-\frac{3\gamma^2}{16(1+\gamma^2)}s_0.\label{evolutioneqn} 
\end{eqnarray}
Here, $\hat{h}s$ denotes the gravitational part of the difference
equation, as given by the left-hand side of \eqref{differenceeqn}.

To deal with \eqref{evolutioneqn}, which is a combination of difference and
differential equations,  
we use standard numerical techniques, implemented in a C source code
written by ourselves. 
For this purpose, we have to discretize the continuous variable $\varphi$.
Since the equations in \eqref{evolutioneqn} are of second order in
$\varphi$, we decompose them into 
a system of first-order equations.
The resulting equations can be solved, for example, with the fourth
order Runge--Kutta algorithm \cite{numerical}.  
Now, we are in a position to calculate the coefficients $s_0$ and
$s_1$ numerically, from which we then get immediately
the functions $s$ in the expansion of the state \eqref{state}.
Our computer code can be found on the link given in \cite{code}.
 
We now briefly describe the steps for the calculation of 
the reduced density matrix. 
As initial data at $\varphi=0$ we have to specify $s_0,\ s_1$, and
their derivatives. 
We choose them to be Gaussians, normalized by the condition ${\rm
  tr}\rho_{\rm red}=1$, except for  
$s_1(\varphi=0)$, which is set equal to zero. 
This provides us with an unentangled state in the beginning of the evolution,
see \eqref{rhored}. 
Since $s_1$ vanishes initially, but not its derivative, the
state will evolve into a mixed state.  
In the long-term evolution, it turns out that $s_1$
will dominate over $s_0$ 
and the state will be pure again, as only the first term in
\eqref{rhored} is left; this, however, happens for times much longer
than the times for which the model is applicable.
 
\begin{figure}[h]
\begin{centering}
\includegraphics[angle=270,width=9cm]{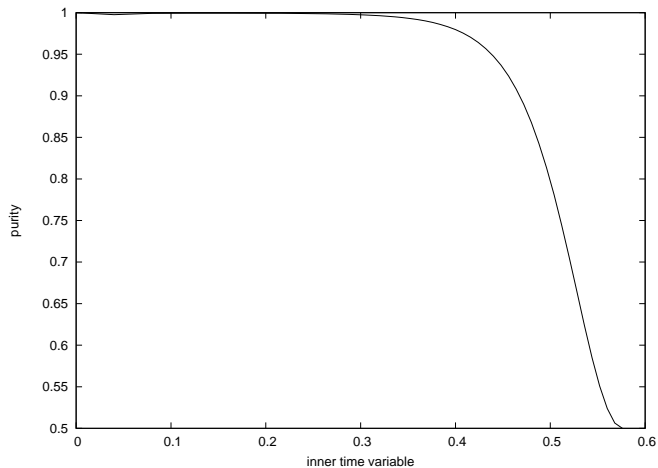} 
\par\end{centering}
 \caption{The purity factor ${\rm tr}\rho_{\rm red}^2$ plotted against
   the inner time variable $\varphi$.} 
 \label{fig:purity}
\end{figure}

The iteration at each time step starts with the calculation of $s_0$
and $s_1$, as described above. 
The evolution preserves the normalization ${\rm tr}\rho_{\rm red}=1$
for the reduced density matrix; small deviations from this condition,
which occur due to numerical errors, are corrected in each time step.
Since the initial state is unentangled,
${\rm tr}\rho_{\rm red}^2$ is initially equal to one. As the inner
time variable increases, the total state becomes entangled, and the
purity factor decreases---the gravitational variables are in a mixed
state, and decoherence becomes more and more efficient. 
The result can be plotted as a function of the inner time variable
$\varphi$, see Fig. \ref{fig:purity}.  

The interesting region is the right half of the diagram, where the
purity factor decreases rapidly.  
In our simulation, the step size $\delta_1$ resp. $\delta_2$
is chosen in such a way that it is large
in the beginning and smaller in the area of interest. 
As a test of the numerical robustness, we checked the algorithm with
different numerical step sizes and different initial data.  
As long as the data are chosen as described above, the resulting
curves look very similar.  
Another test is to choose the simpler Euler algorithm instead of the
 Runge--Kutta procedure. 
The qualitative behaviour is the same for both procedures.

For the calculation, we have to fix the Barbero--Immirzi
parameter $\gamma$ in \eqref{differenceeqn}.  
In principle, $\gamma$ is a free parameter leading to a
one-parameter ambiguity in the quantization.  
In most simulations, we have chosen $\gamma=0.238$, which is motivated by
the desire to get the correct Bekenstein--Hawking entropy for black
holes \cite{meissner04}. However,  
we have found that our results are largely independent of the
specific choice for $\gamma$. 

\section{Conclusion}

In summary, one can say that fermions `measure' the orientation of
the triad in loop quantum cosmology.
In this sense, they provide the universe with a definite
orientation. The fermions act as an environment
similarly to the photons or air molecules that can provide molecules with
a definite shape. Fermions are thus an important ingredient in loop
quantum cosmology. 

If the orientation is fixed by decoherence, it is hard to imagine a
scenario in which the universe changes orientation by going from a pre-
to a post-big bang scenario. One only has different 
branches of the total quantum state with different orientations. These
branches are independent of each other, except possibly for small
universes in which the decohering influence can be neglected,
similarly to small molecules such as ammonia for which the influence
of the environment is too weak to suppress the interferences between
different chiral states. But in this case we expect that the universe is in a
superposition of left and right orientations and that there is no meaning of
a classical transition between the two orientations; a permanent tunnelling
then occurs, similar to the ammonia molecule.

The situation may be compared to a scenario
discussed in string quantum cosmology some time ago \cite{DK97}.
In a semiclassical picture, the standard big bang is preceded in time 
(perhaps through a singularity) by a pre-big bang state.
However, a consistent analysis in the quantum theory shows that wave
packets cannot be continued through the singularity and that pre- and post-
big bang just correspond to independent components of the 
total wave function that decohere from each other. Another example in
quantum cosmology is the decoherence of the `cosmological constant'
(dark energy) by the interaction with metric perturbations
\cite{KQS11}. 

To facilitate the discussion of decoherence as a process in time, we have
introduced a massless scalar field $\varphi$ to serve as an intrinsic
time parameter. 
This, however, implicitly assumes that $\varphi$ is already a
decohered variable. Its 
decoherence can emerge, for example, by the interaction with inhomogeneous
perturbations as discussed in \cite{CK89}. But, again, one would not expect this
decoherence to be efficient for small universes. For this situation, one would
be left with a timeless quantum state for which the traditional
concept of evolution breaks down.

The increase of entropy (decrease of the purity factor) arises because
a special initial state is chosen. This initial state is characterized
by the absence of entanglement between the gravitational and fermionic
degrees of freedom. An initial unentangled state may be at the heart of
the origin of irreversibility in our Universe \cite{Zeh,CK12}. How
this origin can be understood in full loop quantum cosmology
is left for future investigations. 

\vskip 5mm
\noindent
{\bf Acknowledgement}: We thank Martin Bojowald for helpful
discussions and critical comments on an earlier version of our manuscript.



\end{document}